\documentclass[prl,twocolumn,showpacs]{revtex4}
\usepackage{epsfig}

\newcommand{\rem}[1]{}
\newcommand{\beq}{\begin{equation}}
\newcommand{\eeq}{\end{equation}}
\newcommand{\beqa}{\begin{eqnarray}}
\newcommand{\eeqa}{\end{eqnarray}}
\newcommand{\ba}{\begin{array}}
\newcommand{\ea}{\end{array}}

\begin{document}

\title{Self-induced density modulations in the free expansion 
of Bose-Einstein condensates}
\author{Luca Salasnich$^{1}$, Nicola Manini$^{2,3}$, 
Federico Bonelli$^{2}$, Michael Korbman$^{2}$, and Alberto Parola$^{4}$}
\affiliation{
$^{1}$CNISM and CNR-INFM, Unit\`a di Padova, \\
Dipartimento di Fisica ``Galileo Galilei'', Universit\`a di Padova,
Via Marzolo 8, 35131 Padova, Italy \\
$^{2}$Dipartimento di Fisica, Universit\`a di Milano, Via Celoria 16, 
20133 Milano, Italy\\
$^{3}$CNISM, Unit\`a di Milano, Via Celoria 16, 20133 Milano, Italy\\
$^{4}$Dipartimento di Fisica e Matematica, Universit\`a dell'Insubria, 
Via Valleggio 11, 22100 Como, Italy}

\begin{abstract} 
We simulate numerically the free expansion of a repulsive Bose-Einstein
condensate with an initially Gaussian density profile. 
We find a self-similar expansion only for weak inter-atomic repulsion.
In contrast, for strong repulsion we observe the spontaneous formation of a
shock wave at the surface followed by a significant depletion inside the
cloud.
In the expansion, contrary to the case of a classical viscous gas, the
quantum fluid can generate radial rarefaction density waves with several
minima and maxima.
These intriguing nonlinear effects, never observed yet in free-expansion
experiments with ultra-cold alkali-metal atoms, can be detected with the
available setups.
\end{abstract}

\maketitle


The anisotropic free expansion of a gas of $^{87}$Rb atoms was the first
experimental evidence of Bose-Einstein condensation in ultra-cold gases
\cite{anderson}.
The non-ballistic free expansion observed with $^6$Li atoms 
has been saluted as the first signature of superfluid 
behavior in a ultra-cold fermi vapor \cite{ohara}. 
In both cases the atomic quantum gases can be described by the hydrodynamic
equations of superfluids and, because the initial density profile is an
inverted parabola, the free expansion is self-similar
\cite{castin,kagan,ohberg,sala0,Diana06}.

The first theoretical investigations of the free expansion into vacuum of a
classical gas sphere with constant initial density dates back to the 1960's
\cite{molmud,zeldovich}: numerical analysis was needed to analyze in detail
the formation of a depletion at the center and of a shock wave at the
surface \cite{kondrashov,freeman}.
More recently, rarefaction waves have been produced experimentally by the
free expansion of an electron plasma \cite{moody}.

In this Letter we show that the free expansion of a bosonic superfluid into
vacuum displays intriguing nonlinear phenomena.
In particular, by integrating numerically \cite{sala-numerics} the
time-dependent Gross-Pitaevskii equation \cite{book-stringari} we prove
that the expansion of a repulsive Bose-Einstein condensate (BEC) of initial
Gaussian density profile gives rise to self-induced density modulations,
i.e.\ self-induced rarefaction waves.
In addition, we find that the bosonic cloud produces a shock wave at the
surface, that is damped by the quantum pressure of the superfluid.
These nonlinear effects, which can be observed experimentally with
available techniques, are strongly suppressed if a harmonic confinement
along two perpendicular directions is retained and free expansion is
allowed in 1D only.
Observe that the formation of shock and rarefaction waves induced by an
{\it ad hoc} external perturbation has been suggested
\cite{zak,damski,gammal} and observed recently in BECs \cite{hau,cornell}
and also in non-dissipative nonlinear optics \cite{wan}.

We describe the collective motion of the BEC of $N$ atoms in terms of a
complex mean-field wavefunction $\psi({\bf r},t)$
normalized to unity and
such that $\rho({\bf r},t)=N\rho_1({\bf r},t)=N\,|\psi({\bf r},t)|^2$ is
the number-density distribution.
The equation of motion that we assume for $\psi({\bf r},t)$ is 
the time-dependent Gross-Pitaevskii equation (GPE) \cite{book-stringari} 
\beq
\label{GPE}
i\hbar \frac {\partial}{\partial t} \psi =  
\Big[ -\frac{\hbar^2}{2 m} \nabla^2 + U 
+ \!{4\pi\hbar^2 a_{\rm s}\over m} N |\psi|^2 \Big] \, 
\psi \; , 
\eeq
where $U({\bf r},t)$ is a confining potential 
that we assume to vanish at $t\geq 0$ 
(thus allowing free expansion) and $m$ is the atomic mass. 
The nonlinear term represents the inter-atomic interaction 
at a mean-field level, where $a_{\rm s}$ is the 
s-wave scattering length, and we consider the repulsive regime $a_{\rm s} >0$. 

In traditional experiments with ultra-cold alkali-metal atoms
\cite{anderson,ohara} expansion starts from an initial state coinciding
with the ground state of the confined superfluid under the action of a
(often anisotropic) harmonic potential.
For robust interparticle interaction (large number of particles), the
density profile in this intial state resembles closely a negative-curvature
parabola \cite{book-stringari}.
When such density profile is taken as the initial state of a successive
free non-ballistic expansion, to a very good degree of approximation it
expands in a self-similar fashion, maintaining the same shape and only
spreading out and scaling down its height proportionally, until a ballistic
regime is reached when dilution leads to a fully non-interacting regime
\cite{castin,kagan,ohberg,sala0,Diana06}. 

In the present work we discuss the much more exciting phenomena 
observed in the expansion of an interacting condensate starting off 
as a stationary Gaussian 
\beq\label{initialstate}
\psi({\bf r},0) = {1\over (\pi\, \sigma^2)^{3/4} } \, 
\exp\!\left(-\sum_{i=x,y,z} \frac{r_i^2}{2 \sigma_i^2} \right)
,
\eeq
with $\sigma^3 = \sigma_x\sigma_y\sigma_z$. 
Note that a Gaussian profile is readily achieved experimentally by
equilibrating the BEC with a very small scattering length $a_{\rm s}$
obtained by means of the Fano-Feshbach resonace technique
\cite{fano,book-stringari} with a carefully tuned external constant
magnetic field.
$a_{\rm s}$ can then be set to the desired value by a sudden change in the
magnetic field at the time when the harmonic trapping potential 
is switched off.

\begin{figure}
\centerline{\epsfig{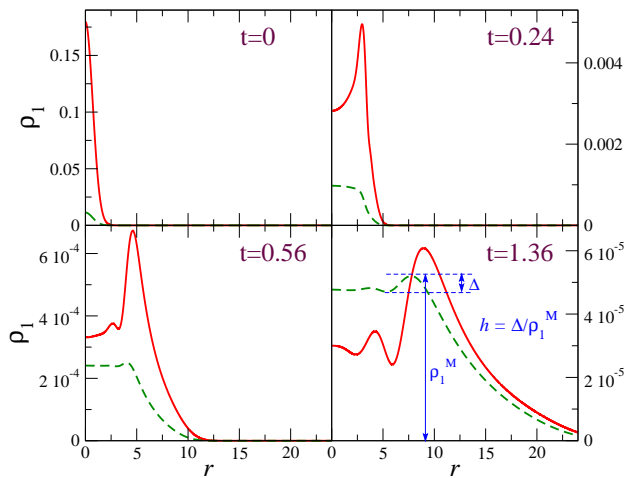}}
\caption{\label{f1:fig} (Color on line). 
Four successive frames of the radial density profile (solid line) for the 
expansion of a strongly interacting condensate, characterized by 
dimensionless interaction parameter $g=2000$. 
The ``opacity'' of the expanding cloud (dashed line) given by the density
$\rho_1({\bf r})$ integrated along lines at a distance $r$ from the center
(rescaled by a factor $0.036$).
All quantities are dimensionless: lengths in units of the initial Gaussian
width $\sigma$ and time in units of $m\sigma^2/\hbar$.}
\end{figure} 

The isotropic case $\sigma_x=\sigma_y=\sigma_z=\sigma$ is conceptually
advantageous, as the expansion can be studied within the GPE model in its
full generality as a function of a single parameter.
Consider rescaling the variables of Eq.~(\ref{GPE}) as follows: ${\bf r}
\to {\bf r}/\sigma$, $t\to t \, \hbar/(m\sigma^2)$, and $\psi \to \psi\,
\sigma^{3/2}$, to produce a dimensionless form
\beq
\label{GPEdl}
i\frac{\partial }{\partial t} \psi = 
\left[ -\frac 12 \nabla^2 + g |\psi|^2 \right] \psi 
\; ,
\eeq
of the equation for the free expansion of a condensate expansion starting
off from a Gaussian initial state of unit width. 
The dimensionless interaction strength
$g = 4\pi\,N\,a_{\rm s}/\sigma$ 
is the one free parameter determining the properties of the free expansion.
A gas of non interacting ($g = 0$) bosons starting from the Gaussian state
expands self similarly according to the formula \cite{robinett}
\beq\label{ideal-psi}
\psi({\bf r},t) = \frac{1}{ \pi^{3/2} (1+t^2)^{3/2}}\,
\exp\!\left(-\frac{r^2 (1+it)}{2(1+t^2)}\right) \, . 
\eeq 
By using an efficient finite-difference Crank-Nicolson algorithm 
\cite{sala-numerics} we have verified that little changes affect 
the expansion as long as the interaction is small. 
$g<1$ tracks the weak-coupling limit where interaction only
accelerates slightly the free expansion, while $g \gg 1$ represents the 
strong-coupling regime, where the mean-field self-interaction term in 
Eq.~(\ref{GPEdl}) dominates the expansion for long enough to produce
substantial nonlinear effects such as those sketched in
Fig.~\ref{f1:fig}. 
In particular we observe the rapid build-up of a sharp expanding spherical
density wave which leaves behind a central region of depleted density. 
New successively formed radial ripples cross this density-depleted region.
Eventually, at very long times, when the overall density has decayed enough
for the nonlinear term in Eq.~(\ref{GPEdl}) to become negligible
everywhere, the expansion recovers a bell-shaped profile. 

\begin{figure}
\centerline{\epsfig{file=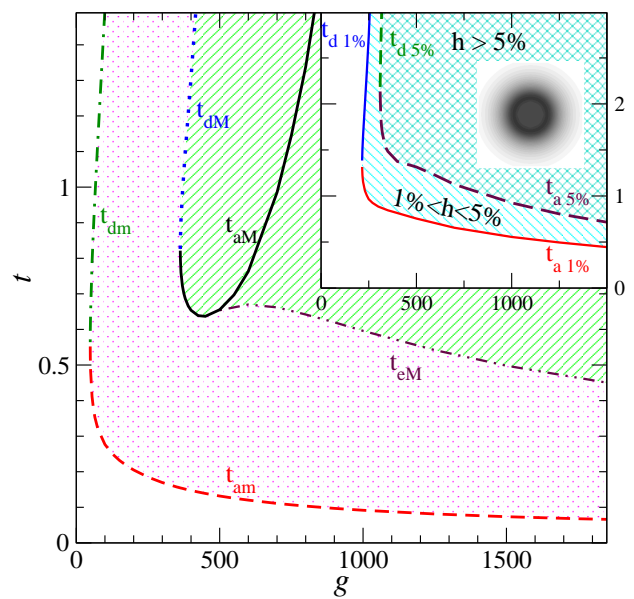,width=82mm,clip=}}
\caption{\label{f2:fig} 
(Color on line). 
Characteristic times of the free expansion as a function of the interaction
strength $g$.
For $g\geq 48.7$, a first minimum in the radial density profile appears at
the center $r=0$ at time $t_{\rm am}$ (dashed) and fills in at $t_{\rm dm}$
(dot-dashed).
For $g\geq 363$, a local maximum re-forms at the center at time $t_{\rm
aM}$ (solid)  and disappears at time $t_{\rm dM}$.
A density maximum appears within the rarefaction region at time $t_{\rm
eM}$ (dot-dot-dashed).
Inset: the time of appearance $t_{\rm a}$ and disappearance $t_{\rm d}$ of
a denser ring (sketched in the small square for $g=1200$, $t=1.8$) of
visibility $h$ (defined in Fig.~\ref{f1:fig}) 1\% (solid) and 5\%
(dashed) in the opacity profile (dashed lines of Fig.~\ref{f1:fig}),
as a function of $g$.
Units as in Fig.~\ref{f1:fig}.
} 
\end{figure}

More quantitatively, for increasing $g$, the rarefaction starts to be
evidenced by a local minimum at the droplet center for $g\gtrsim 48.3$, and
becomes more and more pronounced and long lived for larger interaction 
strength $g$. 
In the highly nonlinear regime, the free expansion of the bosonic cloud
into vacuum develops sequences of radial density waves with minima and
maxima, each starting at a characteristic time and disappearing at a later
time.
Figure \ref{f2:fig} tracks a few early times in this class, as a
function of $g$: more could be defined for larger $g$.
Overall, the time $t_{\rm am}$ of appearance of the first local minimum
reduces slowly as $g$ increases, while the time $t_{\rm dm}$ of
disappearance of all local minima and recovery of a bell-shaped profile
increases rapidly with $g$. 
In between these two characteristic times several traveling density minima
and maxima can be formed depending on the value of $g$.
Generally, the number and density difference of local minima and maxima
increases with the interaction strength.

In practice, the visibility of the central depletion need not be easy to
appreciate by means of total opacity measurements, since such measurements
address the integrated density of a generic linear section crossing the
droplet at a given distance from its center.
However, Fig.~\ref{f1:fig} shows that for suitably strong interaction,
a sensibly higher-opacity outer ring does indeed develop.
The inset of Fig.~\ref{f2:fig} tracks the time range when this inner
optical-density minimum remains lower than the outside denser ring by at
least 1\% and 5\%: it is seen that for strong enough interaction ($g\gtrsim
215$ and $g\gtrsim310$ respectively), the depleted region realizes a limited
but significant visibility, which improves for stronger coupling.
An example of visible opacity ring is displayed in the upper right corner
of Fig.~\ref{f2:fig}.

\begin{figure}
\centerline{\epsfig{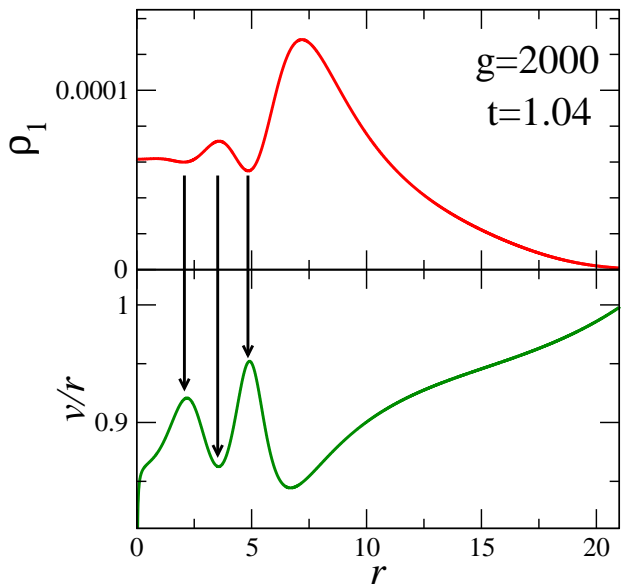}}
\caption{(Color on line). 
Comparison of the density profile $\rho_1(r)=
|\psi(r)|^2$ and of the velocity profile
$v(r)/r$ for the spherically symmetrical expansion of the
strongly-interacting bosonic cloud. 
Units as in Fig.~\ref{f1:fig}.}
\label{f3:fig}
\end{figure}

To characterize the dynamics of the bosonic cloud it is useful to introduce
its local phase velocity, given by
\beq 
{\bf v} = \frac i2 \,
\frac{\psi^* \nabla \psi - \psi \nabla \psi^* }{|\psi|^2} \; . 
\eeq 
This velocity can be written as ${\bf v}({\bf r},t) = \nabla \theta({\bf
r},t)$, where $\theta({\bf r},t)$ is the phase of the macroscopic wave
function $\psi({\bf r},t) = \rho_1({\bf r},t)^{1/2} e^{i\theta({\bf r},t)}$
of the bosonic superfluid.
From Eq.~(\ref{ideal-psi}) one finds immediately the radial phase velocity
of a spherical non-interacting Bose gas ($g=0$): 
\beq 
v = {r\over 2} {t\over 1+t^2} \; . 
\label{ideal-v} 
\eeq
In the interacting case ($g > 0$) the nonlinear term acts as the chemical
potential of a fluid of classical pressure $P=g|\psi|^4/2$ and sound
velocity $c_s=\sqrt{g |\psi|^2}$ \cite{book-stringari}.
For $g\ll 1$ the gas velocity follows Eq.~(\ref{ideal-v}) 
closely, while
for large $g$ deviations from Eq.~(\ref{ideal-v}) are substantial, as
mainly the interaction term, rather than the quantum tendency to
delocalization, drives the expansion.
The ratio $v/r$, a constant as a function of $r$ according to
Eq.~(\ref{ideal-v}) in a non-interacting expansion, shows strong deviations
induced by the nonlinear term $g|\psi|^2$.
In practice, the phase velocity of a strongly interacting BEC approaches
the local sound velocity, with larger densities implying higher velocities.
Accordingly, initially the central part of the strongly repulsive bosonic
superfluid accelerates and propagates faster than the periferic part: the
ensuing mass flow is responsible for the formation of the rarefaction
region inside the cloud, shown in Fig.~\ref{f1:fig}.
The matter flowing quickly out of the central region accumulates near the
profile edge on top of the slower external tail, thus tending to produce a
shock wave \cite{damski,cornell,wan}, 
with a BEC density profile extremely steep at the surface,
approaching a step function: this is illustrated by the
$t\!=\!0.24$ panel of Fig.~\ref{f1:fig}.
This steep wave front survives for a brief period, after which density
oscillations shoot backwards and the surface profile rapidly smoothens its
density gradient.
At this point, the expansion dynamics is strongly affected by these
backward density oscillations, which induce a {\em inverted} relation
between local velocity and density, as Fig.~\ref{f3:fig} illustrates: for
a strongly interacting BEC, at a fixed time $t$ the ratio $v/r$ finds local
maxima (minima) in correspondence to the local minima (maxima) of the
density profile $\rho_1$.
This inversion demonstrates the {\em inward} motion of the density ripples.
These local minima represent the rarefaction waves produced by the surface
step smoothing.
This smoothing is driven by the quantum pressure term \cite{damski}.

\begin{figure}
\centerline{\epsfig{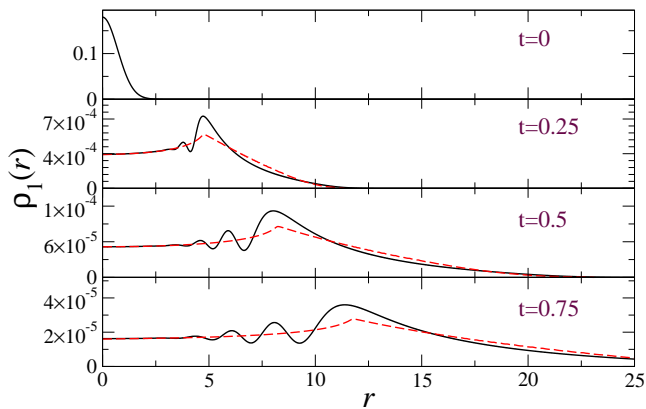}}
\caption{\label{f4:fig}
(Color on line).
Self-depleting radial density profile $\rho_1=\rho/N$ during the expansion:
comparison between the strongly interacting Bose gas (GPE, solid lines) and
the classical gas (NSE, dashed lines), with $g=10^4$ and the same initial
conditions (a unit Gaussian). 
In the NSE, the shear viscosity coefficient $\eta=10^{-5}$.
Units as in Fig.~\ref{f1:fig}.
}
\end{figure}

The quantum pressure $-(|\nabla \psi |^2)/(2 \rho_1)$, which plays a
negligible role in the self-similar non-ballistic expansion \cite{Diana06}
of both Fermi and Bose superfluids with an inverted-parabola initial
profile \cite{ManiniSalasnich05}, becomes relevant in regularizing the
shock-wave singularity, like the dissipative term in classical hydrodynamcs
\cite{cornell}.
Analogous depletion and shock-wave phenomena are indeed observed in the
hydrodynamical expansion of hot classical fluids, and can be simulated
e.g.\ by means of the Navier-Stokes equations (NSE), which depend on the
(dissipative) coefficient of shear viscosity $\eta$ \cite{zeldovich}.
For $\eta=0$ the irrotational (${\bf \nabla}\wedge {\bf v}={\bf 0}$) NSE
reduce to the Euler equations of an ideal (non-viscous) fluid.
By using $P(\rho) = g \rho_1^2/2$ as the equation of state, the Euler
equations are exactly equivalent to the GPE without the quantum pressure
term \cite{book-stringari}.
In Fig.~\ref{f4:fig} we compare the expanding BEC (GPE,
Eq.~(\ref{GPEdl})) and classical gas (NSE), for $g=10000$.
The NSE are solved by means of a Lagrangian finite-difference method
\cite{num-lagr}.
The four successive frames displayed in Fig.~\ref{f4:fig} show that,
while expanding, the interacting BEC and the classical gas produce a
remarkably similar self-depletion of the central region.
On the other hand, the multi-peak rarefaction structures predicted by the
GPE are not reproduced by the NSE.
Thus the novelty of the free expansion of a BEC with respect to that of a
weakly viscous classical fluid, stands mainly in the distinct
large-amplitude rarefaction waves moving inside the depletion region, which
are supported by the nondissipative nature of quantum pressure.

The expansion of the BEC in an anisotropic context is also interesting. 
In particular, we find that a 1D expansion starting from a Gaussian state
produces no shock wave and no central depletion.
This applies both for an idealized 1D BEC, represented by a purely 1D GPE,
and for the 3D expansion of a Gaussian wavepacket to which an harmonic
confining potential is kept along two orthogonal space directions.
Interestingly, if for the initial state of the 3D geometry a spherical
Gaussian (\ref{initialstate}) of width equal to the confining-potential
harmonic length $[\hbar/(m\omega)]^{1/2}$ is taken, the
strongly-interacting BEC shoots out rapidly in all directions, thus
producing a shock wave, depletion, and rarefaction waves around the
cylindrical symmetry axis.
However, the total density, integrated in the directions perpendicular to
the symmetry axis, always retains a bell shape, even for very large
interaction.

In conclusion, we have shown that the free expansion of a Bose-Einstein
condensate reveals novel and interesting nonlinear effects, not achievable
with classical viscous fluids, and which are awaiting experimental
investigations.

The authors thank Paolo Di Trapani, Luciano Reatto and Flavio Toigo for
useful suggestions.

\end{document}